\documentclass[aps,preprint,amsmath,amssymb]{revtex4}
\usepackage{graphicx}
\begin{document}

\title{The effect of electromagnetic properties of neutrinos on the photon-neutrino decoupling temperature}

\author{S.C. \.{I}nan}
\email[]{sceminan@cumhuriyet.edu.tr}
\affiliation{Department of Physics, Cumhuriyet University,
58140, Sivas, Turkey}

\author{M. K\"{o}ksal}
\email[]{mkoksal@cumhuriyet.edu.tr}
\affiliation{Department of Physics, Cumhuriyet University,
58140, Sivas, Turkey}

\begin{abstract}
We examine the impact of electromagnetic properties of neutrinos on the annihilation of relic neutrinos with ultra high energy cosmic neutrinos for the $\nu \bar{\nu}\to \gamma\gamma$ process. For this process, photon-neutrino decoupling temperature is calculated via effective lagrangian model beyond the standard model. We find that photon-neutrino decoupling temperature can be importantly reduced below the QCD phase transition with the model independent analysis defining electromagnetic properties of neutrinos.

\end{abstract}

\maketitle

\section{Introduction}

Neutrinos and photons are the most abundant particles in the
universe. The universe is filled with a sea of relic neutrinos that
decoupled from the rest of the matter within the first few seconds
after the Big Bang. Unlike the relic photons, relic neutrinos have
not been yet observed because of the interactions of their cross
sections with matter are overwhelmingly suppressed. It is very
important to detect relic neutrinos which have played a crucial role
in Big Bang the nucleosynthesis, structure formation and the
evolution of the universe. Nevertheless, some indirect evidences of
the relic neutrinos  may be observed, such as, the UHE neutrinos may
interact with relic neutrinos via the
$\nu_{cosmic}+\overline{\nu}_{relic}\to Z \to nucleons + photons$
reactions occurring on the Z resonance \cite{wei}. If relic
neutrinos do exist, the existence of their mass spectrum may be
reveal with detectors of UHE neutrinos, such as Icecube \cite{ice},
ANITA \cite{ani}, Pierre Auger Observatory \cite{pao}, ANTARES
\cite{ant}.

The $2\rightarrow2$ scattering processes
$\gamma\nu\rightarrow\gamma\nu$,
$\gamma\gamma\rightarrow\nu\bar{\nu}$ and
$\bar{\nu}\nu\rightarrow\gamma\gamma$ have been extensively studied in literature
\cite{abbas, dic1, abb, abbas1, ng}. When the neutrinos are
massless, the $\nu \bar{\nu}\to \gamma\gamma$ process implies a
vanishing cross section from Yang's theorem \cite{cny, mgm} due to
the vector-axial vector nature of the weak coupling. The cross
section of the $\nu \bar{\nu}\to \gamma\gamma$ process can be given
to be of order $G_{F}^{2}\alpha^{2}\omega^{2}(\omega/m_{W})^{4}$
\cite{lev, dic1}. This situation continues to until center of mass
energies $\sqrt{s}\sim 2m_{W}$ where $m_{W}$ is the W boson mass.
The dimension-8 effective lagrangian induced from loop contributions of SM particles
can be given as follows \cite{dic}

\begin{eqnarray}
L^{SM}_{eff}=\frac{i}{32\pi} \frac{g^{2}\alpha}{m^{4}_{W}} A
[\overline{\psi}\gamma_{\nu}(1-\gamma_{5})(\partial^{\mu}\psi)-
(\partial^{\mu}\overline{\psi})\gamma_{\nu}(1-\gamma_{5})\psi]F_{\mu\lambda}F^{\nu\lambda}
\label{l1}
\end{eqnarray}
where $F_{\mu\nu}$ is the electromagnetic field strength tensor, $g$ is the electroweak gauge coupling, $\alpha$ is the fine structure constant and $A$ is given by

\begin{eqnarray}
A=\left[\frac{4}{3} \ln\left(\frac{m^{2}_{W}}{m^{2}_{e}}\right)+1\right].
\end{eqnarray}
It is shown that the equation (\ref{l1}) can be rewritten in the following form \cite{dic},

\begin{eqnarray}
L^{SM}_{eff}=\frac{1}{8\pi} \frac{g^{2}\alpha}{m^{4}_{W}} A
T^{\nu}_{\alpha\beta}T^{\gamma\alpha\beta}
\label{l2}
\end{eqnarray}
where $T^{\nu}_{\alpha\beta}$ and $T^{\gamma\alpha\beta}$ are the stress-energy tensor of the neutrinos and photons which are given by,

\begin{eqnarray}
T^{\nu}_{\alpha\beta}=&&\frac{i}{8}[\overline{\psi}\gamma_{\alpha}(1-\gamma_{5})(\partial_{\beta}\psi)+
\overline{\psi}\gamma_{\beta}(1-\gamma_{5})(\partial_{\alpha}\psi)\nonumber\\
&&-(\partial_{\beta}\overline{\psi})
\gamma_{\alpha}(1-\gamma_{5})\psi-(\partial_{\alpha}\overline{\psi})\gamma_{\beta}(1-\gamma_{5})\psi],
\end{eqnarray}
\begin{eqnarray}
T^{\gamma}_{\alpha\beta}=F_{\alpha\lambda}F_{\beta}^{\lambda}-\frac{1}{4}
g_{\alpha\beta}F_{\lambda\rho}F^{\lambda\rho}.
\end{eqnarray}

 The photons and neutrinos decouple for the $\nu \bar{\nu}\to \gamma\gamma$ process is calculated
 at a temperature $T_c\sim1.6$ GeV, or approximate  one micro second after the Big Bang \cite{abb}.
 If the photon-neutrino interaction can be increased, then decoupling temperature is lowered to the
 QCD phase transition ($\Lambda_{QCD}\sim200$ MeV). Therefore, some remnants of the photons circular
 polarization can possibly be retained in the cosmic microwave background \cite{dic} which can be
 considered as an evidence for the relic neutrino background. Increasing the cross section of $\nu \bar{\nu}\to \gamma\gamma$
 process can be achieved with using models beyond the SM. In this sense, effect of the large extra dimensions
 \cite{dic}, unparticle physics \cite{dut} and excited neutrinos \cite{cem} have been calculated. They have found that the
 photon-neutrino decoupling temperature can be significantly brought down.

 In this study, we have calculated that effect of the electromagnetic properties of neutrinos on the photon-neutrino decoupling temperature for the $\nu \bar{\nu}\to \gamma\gamma$ process.

\section{$\nu \bar{\nu}\to \gamma\gamma$ process including electromagnetic properties of neutrinos}

In the SM, there is no interaction between neutrinos and photons. Besides, minimal extension of the SM with massive neutrinos yields
couplings of $\nu\bar{\nu}\gamma$ and $\nu\bar{\nu}\gamma\gamma$ by means of radiative corrections \cite{Schrock,Marciano,Lynn,Crewther,Feinberg}.
There are a lot of models beyond the SM estimating large enough $\nu\bar{\nu}\gamma$ and $\nu\bar{\nu}\gamma\gamma$ couplings, although minimal extension of the SM give rise to very small couplings. For this reason, it is important to investigate electromagnetic properties of the neutrinos in effective lagrangian methods. Electromagnetic behavior of the neutrinos have significant effects on astrophysics, cosmology and particle physics. In this motivation, we have examined to effect of the Dimension-6 and Dimension-7 effective lagrangians on photon-neutrino decoupling temperature.

\subsection{Dimension-7 Effective Lagrangian}

The dimension-7 effective lagrangian defining
$\nu\bar{\nu}\gamma\gamma$ coupling can be given by
\cite{Nieves,Ghosh,Feinberg,Liu,Gninenko,Larios2}

\begin{eqnarray}
\label{nunuphotonphoton} {\cal
L}=\frac{1}{4\Lambda^3}\bar{\nu}_{i}\left(\alpha^{ij}_{R1} P_R+
\alpha^{ij}_{L1} P_L\right)\nu_{j}\tilde
{F}_{\mu\nu}F^{\mu\nu}+\frac{1}{4\Lambda^3}\bar{\nu}_{i}\left(\alpha^{ij}_{R2}
P_R+ \alpha^{ij}_{L2} P_L\right)\nu_{j} F_{\mu\nu}F^{\mu\nu}
\end{eqnarray}
where $F_ {\mu\nu}$ is the electromagnetic field tensor, $\tilde
{F}_{\mu\nu}=\frac{1}{2}\epsilon_{\mu\nu\alpha\beta}F^{\alpha\beta}$,  $P_{L(R)}=\frac{1}{2}(1\mp\gamma_5)$,
$\alpha^{ij}_{Lk}$ and $\alpha^{ij}_{Rk}$ are dimensionless coupling
constants.
Latest experimental bounds on neutrino-two photon coupling are obtained
from rare decay $Z\to \nu \bar{\nu} \gamma\gamma$ \cite{Larios2} and
the analysis of $\nu_\mu N\to \nu_s N$ conversion \cite{Gninenko}.
The experimental model independent upper limit for  $Z\to \nu \bar{\nu} \gamma\gamma$ decay has been found from the LEP data as follows \cite{Larios2},

\begin{eqnarray}
\label{leplimit} \left[\frac{1 GeV}{\Lambda}\right]^6
\sum_{i,j,k}\left(|\alpha^{ij}_{Rk}|^2+|\alpha^{ij}_{Lk}|^2\right)\leq2.85\times10^{-9}.
\end{eqnarray}
In the external Coulomb field of the nucleus $N$, the model
dependent searches of the Primakoff effect on $\nu_\mu N\to \nu_s N$
conversion founds about two orders of magnitude more restrictive
bound than LEP data. The potential of photon induced reactions at the LHC to probe electromagnetic properties of the
neutrinos has also been studied in the literature for $\Lambda=1$ GeV \cite{sa, sa1}.
It was shown that future experimental
researches at the LHC will place more stringent bounds.
We have used the model independent bound which was
obtained from the LEP data. The contribution of the SM to the $\nu
\bar{\nu}\to \gamma\gamma$ process have been calculated in
Refs.\cite{dic1, abb} with using equation (\ref{l1}). The squared
amplitude for the SM ($|M_{1}|^2$) can be found from this effective
Lagrangian in terms of Mandelstam invariants s and t as below

\begin{eqnarray}
|M_{1}|^2= -16\left(\frac{g^{2}\alpha A}{32\pi M_{W}^{4}}\right)^2t(s^3+2t^3+3ts^2+4t^2s).
\end{eqnarray}

\begin{figure}
\includegraphics{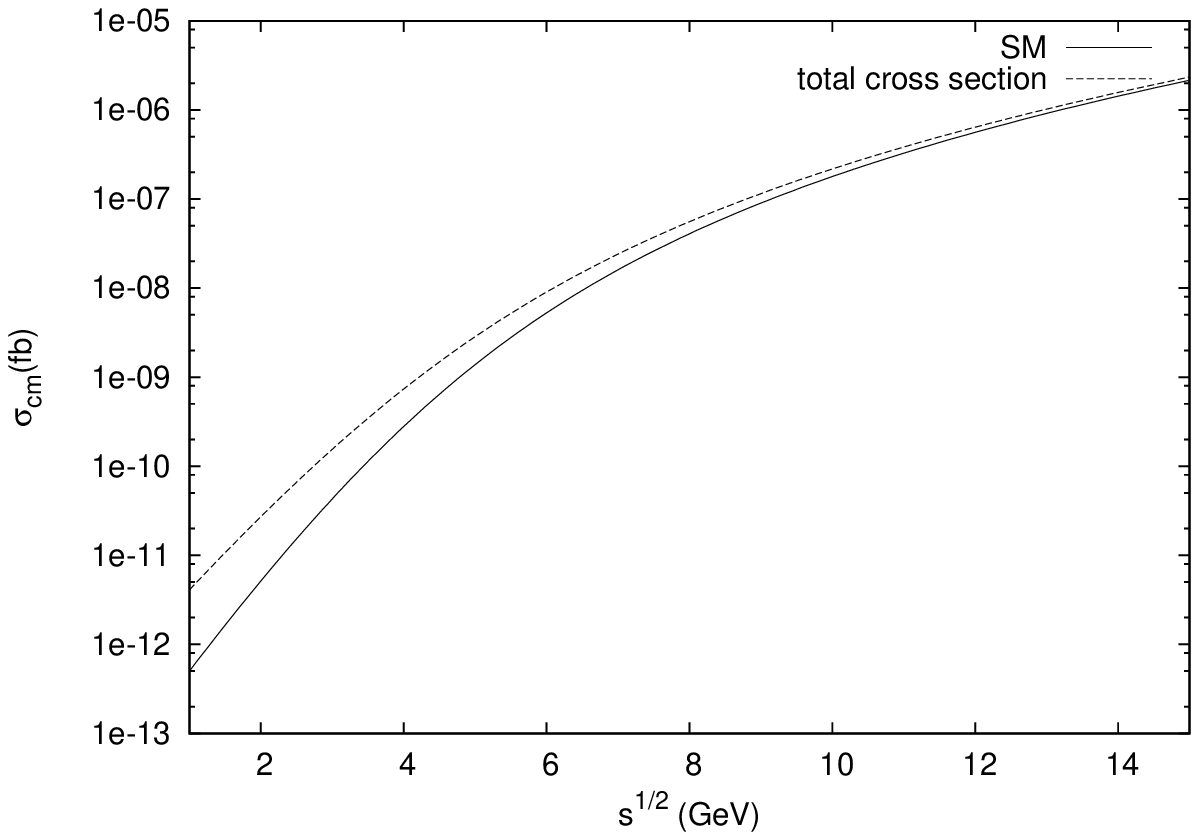}
\caption{The cross sections of $\nu\bar{\nu}\to \gamma\gamma$
process as a function center of mass energy $s^{1/2}$ when $\beta^{2}$
parameter is taken to be $2.89\times10^{-9}$. \label{fig1}}
\end{figure}

\begin{figure}
\includegraphics{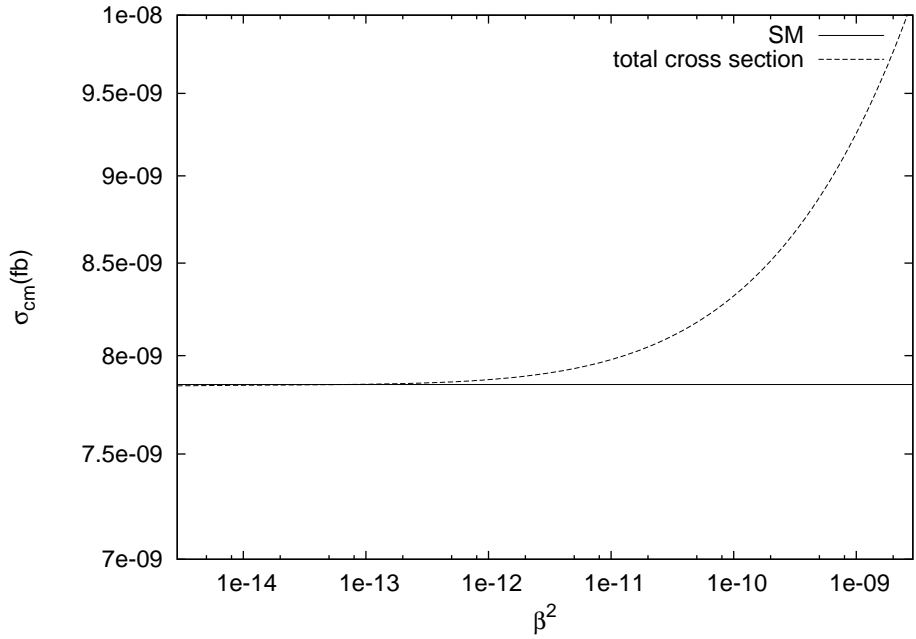}
\caption{The SM and total cross sections of $\nu\bar{\nu}\to
\gamma\gamma$ process as a function $\beta^{2}$ for $s^{1/2}=5$
GeV.
\label{fig2}}
\end{figure}

The new physics contribution with using
equation (\ref{nunuphotonphoton}) comes from t and u channels
diagrams for the $\nu \bar{\nu}\to \gamma\gamma$ process. The
polarization summed amplitude dimension-7 effective interaction
square $(|M_{2}|^2)$ is given below,

\begin{eqnarray}
\label{amplitude1} |M_{2}|^2=\frac{s^3}{8\Lambda^{6}}
\sum_{i,j,k}\left(|\alpha^{ij}_{Rk}|^2+|\alpha^{ij}_{Lk}|^2\right).
\end{eqnarray}
It has been obtained that there is no contribution from the interference term of the SM and dimension-7 effective interaction to the $\nu \bar{\nu}\to \gamma\gamma$ scattering. The reason is that the SM interaction contains neutrinos of opposite helicity, dimension-7 effective interaction contain neutrinos of the same helicity. Hence, the total squared amplitude can be found,

\begin{eqnarray}
|M|^2=|M_{1}|^2+|M_{2}|^2.
\end{eqnarray}
For $\nu \bar{\nu}\to \gamma\gamma$ process, the differential cross
section can be obtained by using

\begin{eqnarray}
\frac{d\sigma}{dz}=\frac{1}{2!}\frac{1}{32\pi s}|M|^{2}.
\label{dcs}
\end{eqnarray}
Therefore, we get the total cross section ($\sigma_{cm}$) as follows,

\begin{eqnarray}
\sigma_{\nu \bar{\nu}\to
\gamma\gamma}=\frac{s^{3}}{20\pi}\left(\frac{g^{2}\alpha A}{32\pi M_{W}^{4}}\right)^2
+\frac{s^{2}}{256\pi\Lambda^{6}}\sum_{i,j,k}\left(|\alpha^{ij}_{Rk}|^2+|\alpha^{ij}_{Lk}|\right).
\end{eqnarray}
We have showed as a function of the center of mass energy $\sqrt{s}$
for both the $SM$ and total cross sections in fig. (\ref{fig1}).
During numeric analysis we have assumed to $\Lambda=1$ GeV to
compare our results with current experimental LEP limit. In this
figure, $\beta^2=\sum_{i,j,k}\left(|\alpha^{ij}_{Rk}|^2+|\alpha^{ij}_{Lk}|\right)$
is taken to be $2.89\times10^{-9}$ which is current experimental LEP
bound. It has been shown that deviation from the SM increases as the
$\sqrt{s}$ decreases. Also, Fig. (\ref{fig2}) shows that the SM and total
cross sections via the $\beta^2$ for $\sqrt{s}=5$ GeV. The total
cross section is nearly the same as the SM at $\beta^2\sim10^{-13}$.
This value almost $10^4$ times larger than the current experimental
LEP limit. Specific values of the  $\beta^2$ and $\sqrt{s}$ total
cross section can be easily discerned from the SM cross section.
Therefore, dimension-7 effective interaction can effect to the
photon-neutrino decoupling temperature.

The temperature at which the $\nu \bar{\nu}\to \gamma\gamma$ process ceases to take place can be found from the reaction rate per unit volume,

\begin{eqnarray}
\rho=\frac{1}{(2\pi)^{6}}\int\frac{d^3\vec{p_1}}{\exp(E_1/T)+1}
\int\frac{d^3\vec{p_2}}{\exp(E_2/T)+1}\sigma|\vec{\upsilon}|.
\label{roo}
\end{eqnarray}
 where $\vec{p_1}$ and $\vec{p_2}$ are the momentums of the neutrinos, $E_1$ and $E_2$  are the energies of the neutrinos, T is the temperature, $|\vec{\upsilon}|$ is the flux. The  $\sigma|\vec{\upsilon}|$ can be obtained in terms of $\sigma_{cm}$ in the center of mass frame by using of invariance of $\sigma|\vec{\upsilon}| E_1E_2$

\begin{eqnarray}
\sigma|\vec{\upsilon}|=\frac{\sigma_{cm}s}{2 E_{1}E_{2}}
\end{eqnarray}

\begin{eqnarray}
\sigma|\vec{\upsilon}|=\frac{s^{4}}{40\pi E_1E_2}\left(\frac{g^{2}\alpha A}{32\pi M_{W}^{4}}\right)^2+
\frac{s^{3}\beta^2}{512\pi E_{1}E_{2}}
\end{eqnarray}
where $s=2 E_{1} E_{2}(1-\cos\theta_{12})$ and $\theta_{12}$ is the
angle between $\vec{p_1}$ and $\vec{p_2}$. Then the reaction rate
per unit volume can be obtained as follows,

\begin{eqnarray}
\rho=&&\frac{g^{4}\alpha^{2}A^{2}}{25 (2\pi)^{7}m^{8}_{W}}
T^{12}\int^{\infty}_{0}\frac{x^{5}
dx}{e^{x}+1}\int^{\infty}_{0}\frac{y^{5} d y}{e^{y}+1}+
\frac{\beta^2}{4(2\pi)^{5}}T^{10}\int^{\infty}_{0}\frac{x^{4}
dx}{e^{x}+1}\int^{\infty}_{0}\frac{y^{4} d y}{e^{y}+1}
\label{roo1}
\end{eqnarray}
where $x=E_1/T$ and $y=E_2/T$. The integration is easily written by

\begin{eqnarray}
\rho=&&\frac{g^{4}\alpha^{2}A^{2}}{25 (2\pi)^{7}m^{8}_{W}} T^{12}
\left[\frac{31}{32}\Gamma (6) \zeta(6) \right]^{2}+\frac{\beta^2}{4(2\pi)^{5}}T^{10}
\left[\frac{15}{16}\Gamma (5) \zeta(5) \right]^{2}
\end{eqnarray}
where $\zeta(x)$ is the Riemann Zeta function. At temperature $T$, the interaction rate $R$ can be found by dividing $\rho$ by the neutrino density $n_{\nu}=3\zeta(3)T^{3}/4\pi^{2}$,

\begin{eqnarray}
R=2.30\times10^{4}\left (\frac{T}{GeV}\right)^{9}+
2.31\times10^{23}\beta^{2} \left (\frac{T}{GeV}\right)^{7} sec^{-1}.
\label{rr}
\end{eqnarray}
Multiplying equation (\ref{rr}) by the age of the universe,

\begin{eqnarray}
t=1.48\times10^{-6} \left(\frac{T}{GeV}\right)^{-2}
\end{eqnarray}
at least one interaction to occur is $R t= 1$. As a result, the decoupling temperature can be found with solution of the following equation,

\begin{eqnarray}
3.40\times10^{-2}\left
(\frac{T}{GeV}\right)^{7}+3.42\times10^{17}\beta^{2}\left
(\frac{T}{GeV}\right)^{5}=1.
\end{eqnarray}
In Fig. (\ref{fig3}) we have plotted the solution of the this equation for different values of the $\beta^2$. Here, current experimental LEP bound have taken to be maximum value of the $\beta^2$.

\subsection{Dimension-6 Effective Lagrangian}

The Dimension-6 effective lagrangian  for non-standard $\nu\bar{\nu}\gamma$ interaction \cite{Larios1,Maya,Larios2,Bell}
is given by

\begin{eqnarray}
L=\frac{1}{2}\mu_{ij}\bar{\nu}_{i}\sigma_{\mu\nu}\nu_{j}F^{\mu\nu}
\label{lanm}
\end{eqnarray}
here $\mu_{ii}$ is the magnetic moment of $\nu_i$ and $\mu_{ij}$
$(i\neq j)$ is the transition magnetic moment. In equation (\ref{lanm}), new physics energy scale $\Lambda$ is absorbed
in the definition of $\mu_{ii}$. We will examine $\nu\bar{\nu}\gamma$ interaction on the $\nu \bar{\nu}\to \gamma\gamma$ process
assuming neutrino magnetic moment matrix is virtually flavor diagonal and only one of the matrix elements is different from zero.
Also, the standard relic neutrinos is considered to comprise of the three active neutrinos of the SM. Current experimental bounds on neutrino magnetic moment are
stringent. The most sensitive bounds from neutrino-electron
scattering experiments with reactor neutrinos are at the order of
$10^{-11}\mu_B$ \cite{Li,Wong1,Wong2,Daraktchieva}. Bounds derived
from solar neutrinos are at the same order of magnitude
\cite{Arpesella}. Bounds on magnetic moment can also be derived from
energy loss of astrophysical objects. These give about an order of
magnitude more restrictive bounds than reactor and solar neutrino
probes
\cite{Raffelt,Castellani,Catelan,Ayala,Barbieri,Lattimer,Heger}.

The polarization summed amplitude square for the $\nu \bar{\nu}\to \gamma\gamma$ process is given by the following equation,

\begin{eqnarray}
|M|^2=-16\left(\frac{g^{2}\alpha A}{32\pi M_{W}^{4}}\right)^2t(s^3+2t^3+3ts^2+4t^2s)
+16\mu^{4}tu+32\mu^{2}t u s \left(\frac{g^{2}\alpha A}{32\pi M_{W}^{4}}\right).
\end{eqnarray}
Then the total cross section for the $\nu \bar{\nu}\to \gamma\gamma$ process can be obtained as follows,

\begin{eqnarray}
\sigma_{\nu \bar{\nu}\to \gamma\gamma}=&&\int_{-1}^1 dz \frac{d\sigma}{dz} \nonumber \\
=&&\frac{s^3}{20\pi}\left(\frac{g^{2}\alpha A}{32\pi M_{W}^{4}}\right)^2 \nonumber \\
+&&\frac{\mu^{2}s}{12\pi}\left(\mu^{2}+2s\left(\frac{g^{2}\alpha A}{32\pi M_{W}^{4}}\right)\right).
\end{eqnarray}
We have calculated the total cross section with using experimental limits of the neutrino magnetic moments ($\mu_{\nu_{i}}, i=e,\mu,\tau $) for the $\nu \bar{\nu}\to \gamma\gamma$ process. These bounds are $\mu_e=3.2\times10^{-11}\mu_B$, $\mu_\mu=6.8\times10^{-10}\mu_B$ and $\mu_\tau=3.9\times10^{-7}\mu_B$ \cite{data}.  It has been seen that there are barely contribution from neutrino magnetic moments to the SM cross section of this process and we have not shown results in here. Therefore, this effective interaction must not reduce to photon-neutrino decoupling temperature significantly. This result can be seen with using same procedure as above. Then, the $\rho$ and $R$ are calculated by,

\begin{eqnarray}
\rho=&&\frac{g^{4}\alpha^{2}A^{2}}{25 (2\pi)^{7}m^{8}_{W}} T^{12}
\left[\frac{31}{32}\Gamma (6) \zeta(6) \right]^{2}+\nonumber\\ &&
\frac{\mu^{2}}{18\pi^{5}}\left(6\left(\frac{g^{2}\alpha A}{32\pi M_{W}^{4}}\right)
T^{10}\left[\frac{15}{16}\Gamma (5) \zeta(5) \right]^{2}+\mu^{2}T^{8}\left[\frac{7}{8}\Gamma (4) \zeta(4) \right]^{2}\right),
\end{eqnarray}

\begin{eqnarray}
R=2.30\times10^{4}\left
(\frac{T}{GeV}\right)^{9}+
8.85\times10^{13}\mu^{2} \left
(\frac{T}{GeV}\right)^{7}+9.71\times10^{22}\mu^{4}
\left (\frac{T}{GeV}\right)^{5} sec^{-1}.
\label{rr1}
\end{eqnarray}
The solution of the following equation gives the decoupling temperature for photon-neutrino coupling,

\begin{eqnarray}
3.40\times10^{-2}\left
(\frac{T}{GeV}\right)^{7}+1.31\times10^{7}\mu^{2}\left
(\frac{T}{GeV}\right)^{5}+1.44\times10^{17}\mu^{4}\left
(\frac{T}{GeV}\right)^{3}=1.
\end{eqnarray}
From this equation, we have found that the photon-neutrino decoupling temperature almost same the SM ($T_c\sim1.6$ GeV) when we used the experimental bounds on neutrino magnetic moments as we expected.

\section{Conclusion}

\begin{figure}
\includegraphics{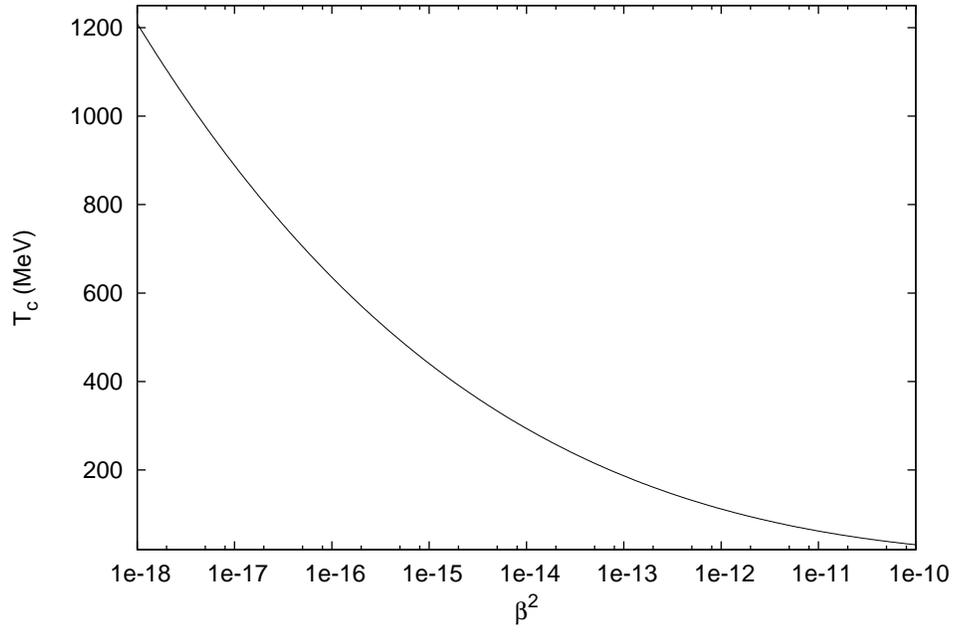}
\caption{The decoupling temperature $T_{c}$ as a function of
$\beta^{2}$.
\label{fig3}}
\end{figure}

If neutrino-photon decoupling temperature can be decreased to below
the QCD phase transition ($\Lambda_{QCD}\sim200$ MeV), this could be
an evidence for the relic neutrino background. Because some remnant
the circular polarization could possibly be sustained in the cosmic
microwave background. For reducing decoupling temperature, the total
cross section of the  photon-neutrino process should be increased.
This can be done with contribution of new effective interactions. In
this motivation, we have examined the effect of electromagnetic
properties of the neutrinos on the photon-neutrino decoupling
temperature with interaction of relic neutrinos with UHE cosmic
neutrinos via the $\nu\bar{\nu}\to \gamma\gamma$ process. First, we
have investigated to dimension-7 effective interaction effect on
$\nu\bar{\nu}\to \gamma\gamma$ process. It is found that this
effective interaction contribution to total cross section of the
$\nu\bar{\nu}\to \gamma\gamma$ process is significant depending on
the $\beta^2$. Therefore, photon-neutrino decoupling temperature can
be reduced below the $\Lambda_{QCD}$ as seen from the
Fig.\ref{fig3}. On the other hand, even if $\beta^2$  is eight order
of magnitude smaller than current experimental bound, this effective
interaction can reduce to $T_c$ below the obtained value of the SM.

Second, we have examined to dimension-6 effective interaction impact
on $\nu\bar{\nu}\to \gamma\gamma$ process. This effective
interaction describes neutrino magnetic moment. Current experimental
bounds on neutrino magnetic moment are stringent. Therefore, the
contribution of the this effective interaction very tiny on
the SM cross section $\nu\bar{\nu}\to \gamma\gamma$. Hence, the
photon-neutrino cross section decoupling temperature is not almost changed.

Consequently, we have shown that dimension-7 effective interaction
can permit of reduced the decoupling temperature for the
$\nu\bar{\nu}\to \gamma\gamma$ process.



\end{document}